# Complex biphase nature of the superconducting dome of the FeSe phase diagram

V. Svitlyk[1]*, M. Raba[2,3], V. Dmitriev[4], P. Rodière[2,3], P. Toulemonde[2,3], D. Chernyshov[4], M. Mezouar[1]

[1]ID27 High Pressure Beamline, European Synchrotron Radiation Facility, 38000, Grenoble, France

[2]Université Grenoble Alpes, Institut NEEL, F-38000 Grenoble, France

[3]CNRS, Institut NEEL, F-38000 Grenoble, France

[4]Swiss-Norwegian Beamlines, European Synchrotron Radiation Facility, 38000, Grenoble, France

*svitlyk@esrf.fr

## Abstract

Single crystal synchrotron X-ray diffraction as a function of temperature and pressure has revealed a complex biphase mixture in superconducting FeSe. Based on our experimental results we construct a phase diagram where structural behavior and superconducting properties of FeSe are found to be correlated. We show that below 6 GPa, where pressure promotes the superconducting critical temperature, the FeSe structure is composed of 2D layers of edge-shared $FeSe_4$ tetrahedra, while above 6 GPa the superconductivity is strongly suppressed on formation of a new orthorhombic polymorph characterized by a 3D network of face sharing $FeSe_6$ octahedra. Therefore changes in topology and connectivity of the FeSe structure are found to be detrimental for superconductivity to exist. This previously controversial crystal structure of the high pressure polymorph of FeSe was also unambiguously determined. High pressure FeSe adopts an orthorhombic MnP-type structure (*Pnma*) which corresponds to a slightly distorted hexagonal NiAs-type arrangement ($P6_3/mmc$). The structural transformation from the low- to high-pressure FeSe polymorph is first order in nature and is manifested as antiparallel displacements within the Fe and Se sublattices.

## Introduction

In cuprate and iron based superconductors the binding of electrons into Cooper pairs is mediated by spin-spin interactions. In the magnetic pairing model the superconducting critical temperature, $T_c$, is enhanced in the case of two-dimensional (2D) structures that promote coupling between conduction electrons and spin fluctuations[1,2,3,4]. However, thermodynamics can spoil a game: even simple solids may show a variety of crystal structures coexisting under the same thermodynamic conditions. Different polymorphs of the same material may vary topologically and, therefore, a modification from a 2D layered structure to an environment with 3D connectivity of magnetic atoms may strongly affect the superconducting properties. In this work we show that the high pressure dependence of the critical temperature of FeSe is indeed correlated with a change of dimensionality of the iron lattice.

The discovery of superconductivity in iron arsenides[5] in 2008, and shortly after in $\beta$-FeSe[6] and its intercalated derivatives,[7,8,9] marked the beginning of the 'iron age' of superconductivity. $\beta$-FeSe itself crystallizes in a relatively simple tetragonal PbO-type structure (*P*4/*nmm* space group, Fig. 1, (a)) and is superconducting below $T_c$ = 8–9 K. The $T_c$ value can be dramatically increased by intercalating the structure[10] ($T_c$ > 40 K) or by applying high pressure (HP)[11] ($T_c$ ~ 37 K). Moreover, monolayers of the FeSe films were found to become superconducting at 65 K[12] and even above 100 K in other studies[13].

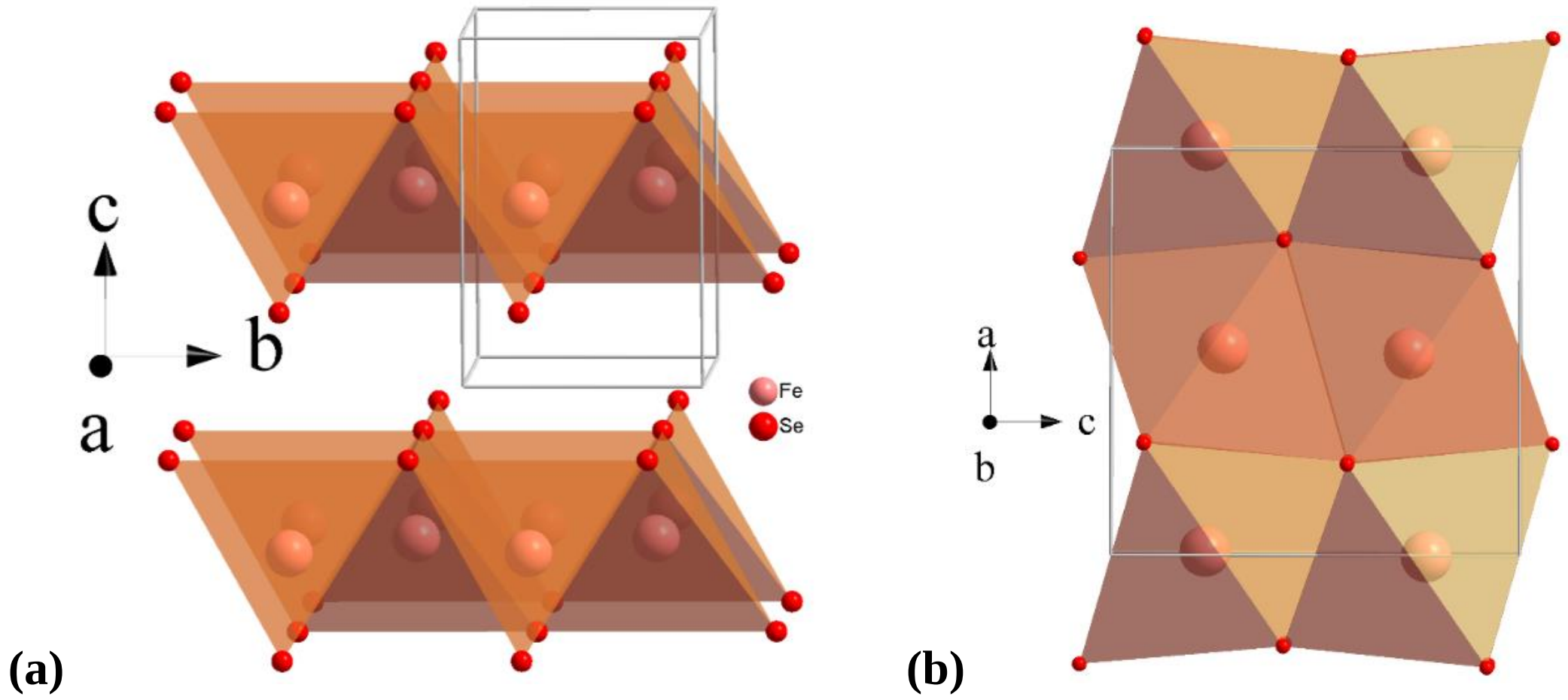

**Figure 1. Low pressure (LP) and high pressure (HP) structures of FeSe** (a) The tetragonal (*P*4/*nmm*) PbO-type $\beta$ modification of FeSe is composed of layers of edge-shared $FeSe_4$ tetrahedra (b) the MnP-type (*Pnma*) FeSe phase at high pressure is composed of chains of face shared $FeSe_6$ octahedra.

Although the $\beta$-FeSe phase has a relatively simple crystallographic structure (Fig. 1(a)) and features no long-range magnetic order at low temperature (LT), its behavior as a function of temperature ($T$) and pressure ($P$) is not well understood. Below ~1 GPa and around $T_s$ = 90 K[14] it undergoes a transformation to the orthorhombic (*Cmma*) nematic-type phase[15,16] with a loss of tetragonal symmetry. At pressures above 1 GPa an antiferromagnetic (AFM) phase has been detected by muon spectroscopy[17,18]. This was tracked recently also in transport measurements on single crystals[19] that show that the AFM phase exists up to 6 GPa[20] in coexistence with the superconducting phase. This AFM phase seems to retain the orthorhombic distortion of the *Cmma* structure[21]. Such a magnetic phase appears to compete with superconductivity and affects its pressure dependence of $T_c$. The $T_c$ variation shows a double dome feature with a first maximum of 13 K at 0.8–1 GPa, followed by a local minimum of 10 K at 1.2–1.5 GPa[22,23], then an important enhancement[24] up to 34–37 K for $P$ = 6–9 GPa[11,25,26,27] and a further progressive decrease at higher pressure due to a structural transition at 10–12 GPa towards a HP, presumably non superconducting, phase[26,27]. Two symmetries were proposed for the HP phase, hexagonal[11] or orthorhombic[26], based on powder X-ray diffraction (XRD) data. However, the unambiguous determination of the crystallographic structure of this HP phase was lacking mainly due to the limitation of the X-ray powder diffraction method in resolving 1D diffraction signals from the disordered (structurally or microstructurally) and likely strained FeSe phase[25,26,27]. Nevertheless, the orthorhombic *Pnma* (*Pbnm)* MnP-type model, suggested experimentally, was supported by the DFT calculations[28,29].

Evidently, for this topical material a high resolution crystallographic study is critical to correctly understand the pressure dependence of its macroscopic response, *i.e.* its $T_c$ variation. In that sense, single crystal diffraction is well adapted to help to unveil structural details of the HP FeSe polymorph. Establishing the correct structure-property relations could help our understanding to further enhance the superconducting properties of related iron selenides through targeted structural modifications and optimizations.

Here we present high-quality single crystal XRD data collected with synchrotron radiation under hydrostatic conditions in broad range of temperatures (from room temperature (RT) down to 20 K) and pressures (in the 0.2–19 GPa range), including in the superconducting domain. The work presented allows us to revisit the phase diagram of FeSe, *i.e.* unambiguously solve the structure of the HP polymorph, explain the structural mechanism linking this HP phase with the low pressure (LP) phase, and consequently uncover the complex biphase nature of the superconducting dome.

## Results and discussion

### Low pressure ($P$ < 6 GPa) behavior of FeSe

The main focus of this report is the HP phase of FeSe that appears above 6 GPa. Nevertheless, we have also studied the transformation of the orthorhombic (*Cmma*) LT-LP phase under moderate pressure. Our analysis shows that at 20 K the ambient pressure orthorhombic distortion of the lattice remains nearly constant up to 6 GPa, *i.e.* in the AFM phase which gradually appears in the 1–2 GPa range, in agreement with the recent report of Kothapalli *et al.* (ref. 21; see XRD patterns on Fig. 1S and *a*-*b* axis pressure dependence on Fig. 2S(a)). Interestingly, we detect an anomaly in the *c*-axis and lattice volume at 1.9(2) GPa (see Fig. 2S(b)), in agreement with the extrapolated curve of $T_s$ that characterizes the nematic transition. This anomaly at 1.9 GPa corresponds also to the onset of the strong enhancement of $T_c$ and is probably related to changes in hybridization between Se and Fe orbitals, as previously evidenced by the Se height and Fe-Se bond length changes found at 2 GPa and 16 K in the powder XRD study of Margadonna *et al.* [see Fig. 4(a) and (c) in ref. 27].

At 50 K, however, the *Cmma* orthorhombic distortion disappears at 1.0(1) GPa, and the ambient condition *P*4/*nmm* tetragonal phase is recovered, in agreement with the extrapolated pressure dependence of $T_s$ obtained from resistivity measurements using the same pressure transmitting medium (PTM, helium)[30]. The tetragonal phase is stable in the 1–2 GPa range. At higher pressures the lattice distorts again, as evidenced by the broadening of the (04l)-(40l) Bragg peaks (Fig. 1S and Fig. 3S). This corresponds to the onset of the AFM phase detected by muon spectroscopy[18], but not exactly to the onset obtained from other probes like transport measurements[19,20] or Mössbauer spectroscopy[21] since these techniques present different sensitivities. Our conclusions on this LP part of the *P*-*T* phase diagram, as established from the experimental lattice distortions (Figs. 2S and 3S), are summarized in Fig. 2, with a comparison to the literature. The comparison of orthorhombicity (defined as $\delta = (b - a)/(b + a)$ parameters ratio) of the ortho-I *Cmma* phase between 20 K and 50 K datasets is shown in the Supplementary information (Fig. 4S). While in the superconducting phase (coexisting with the AFM phase), i.e.

for dataset at 20 K and for pressure above 1.7-2 GPa, orthorhombicity remains nearly constant (ca. 1.7-1.9·$10^{-3}$, value also found at 1.7 and 3.1 GPa by Kothapalli et al., ref. 21), in the AFM phase, i.e. for dataset at 50 K and for pressure above 2 GPa, orthorhombicity continuously increases and reaches doubled values (ca. 3.3·$10^{-3}$) at the onset of the transition towards the HP *Pnma* phase. This illustrates the competition between superconductivity and magnetism: superconductivity tends to reduce the lattice distortion while magnetism increases it.

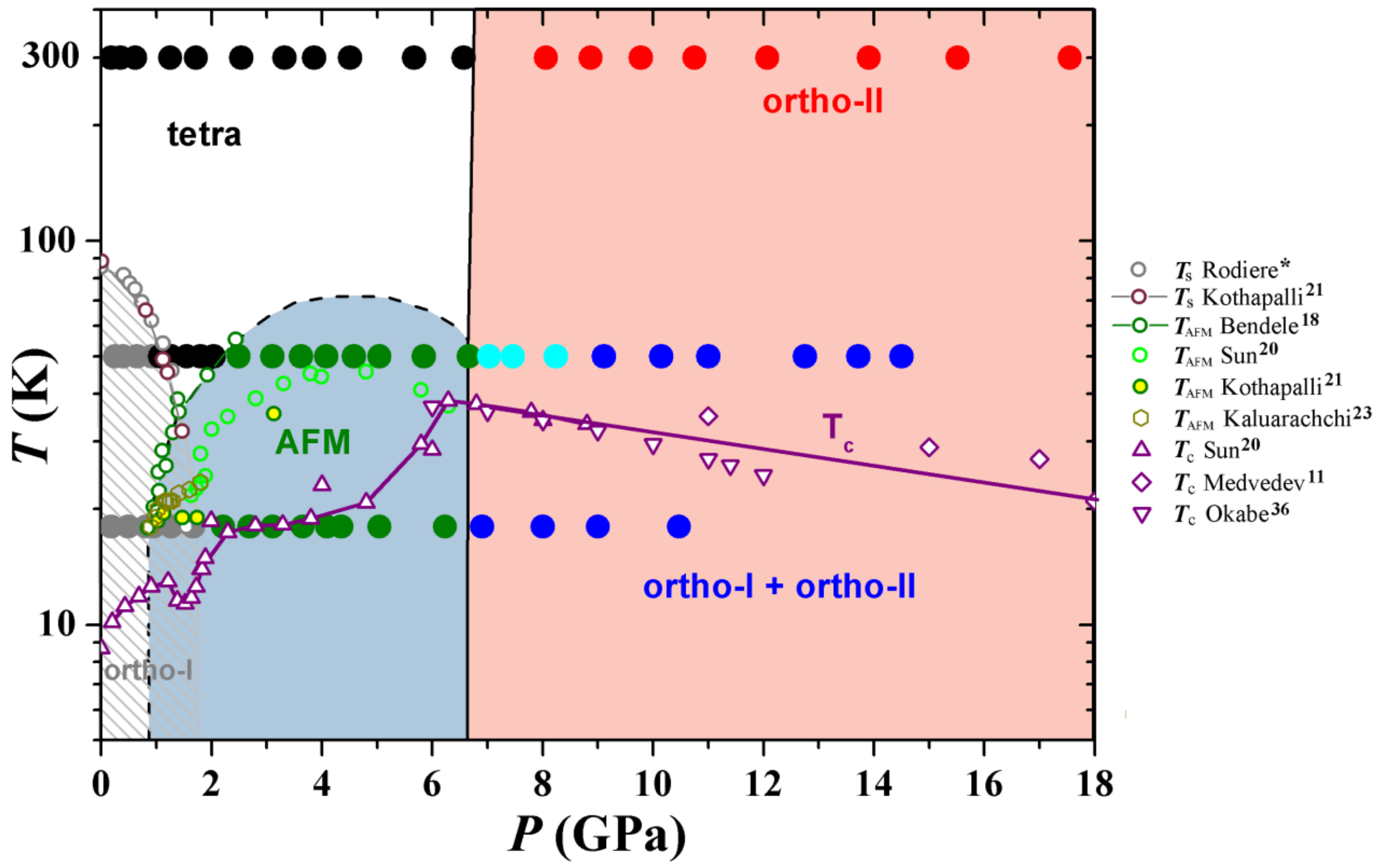

**Figure 2. *T-P* phase diagram of FeSe**. Large filled circles at the 20, 50 and 300 K isotherms correspond to the experimental data obtained in this work (black – tetragonal phase, grey – "ortho-I" phase, green – AFM phase, red – "ortho-II" phase, blue – a mixture of "ortho-I" and "ortho-II" phases, cyan – possible onset of the formation of the "ortho-II" phase; * - our unpublished data based on transport measurements below 1.3 GPa.

## HP structure and the mechanism of LP to HP phase transformation

The LP-HP transition of FeSe was previously studied on powder samples. However, experimental X-ray diffraction patterns of the HP FeSe modification degrade rapidly with pressure (see below) that resulted in numerous controversies in the literature. The formation of the hexagonal (NiAs-type) FeSe form was suggested at pressures between 8.5 and 12 GPa at RT (He gas as PTM[25]) and above 9 GPa at 16 K (Daphne oil as PTM[27]). The HP FeSe was identified as orthorhombic with a help of powder diffraction, but with a higher transition pressure of 12 GPa (He gas as a PTM[26]). Apparently, the observed transition pressures were influenced by the PTMs and by the pressure increase rate. Indeed the latter plays a significant role on the observed behavior for first-order structural transformations in Cs-intercalated FeSe[31].

In our single crystal XRD study we observe clearly the transformation of the LP phase to the HP phase at RT between 6.7 and 8.1 GPa where only the HP form is present. Our structural model of the HP FeSe polymorph was unambiguously obtained from the integrated intensities of the single crystal data collected at 8.1 GPa and RT. It is found to correspond to the orthorhombic MnP-type structure[32] (Table 1, Fig. 1(b)) and features chains of face sharing $FeSe_6$ octahedra along the *a* axis and, from the first point of view, is not directly symmetry related to the parent LP tetragonal *P*4/*nmm* FeSe phase. The experimental reciprocal *h*0*l* layer of the HP FeSe phase is shown on Fig. 3 and it is consistent with a *Pnma* symmetry.

**Table 1.** Atomic parameters of HP FeSe from single crystal diffraction at 8.1 GPa and 300 K. The structure was solved in *Pnma* (equivalent to *Pbnm*), $a$ = 5.912(4), $b$ = 3.457(10), $c$ = 5.952(4) Å, $z$ = 4, $R_1$ = 0.12 with anisotropic refinement of atomic displacement parameters (ADPs).

| Atom | *x* | *y* | *z* | $U_{eq}$, Å$^2$ |
|---|---|---|---|---|
| Fe | 0.0069(8) | 1/4 | 0.2259(9) | 0.046(3) |
| Se | 0.2237(8) | 1/4 | 0.5786(7) | 0.083(4) |

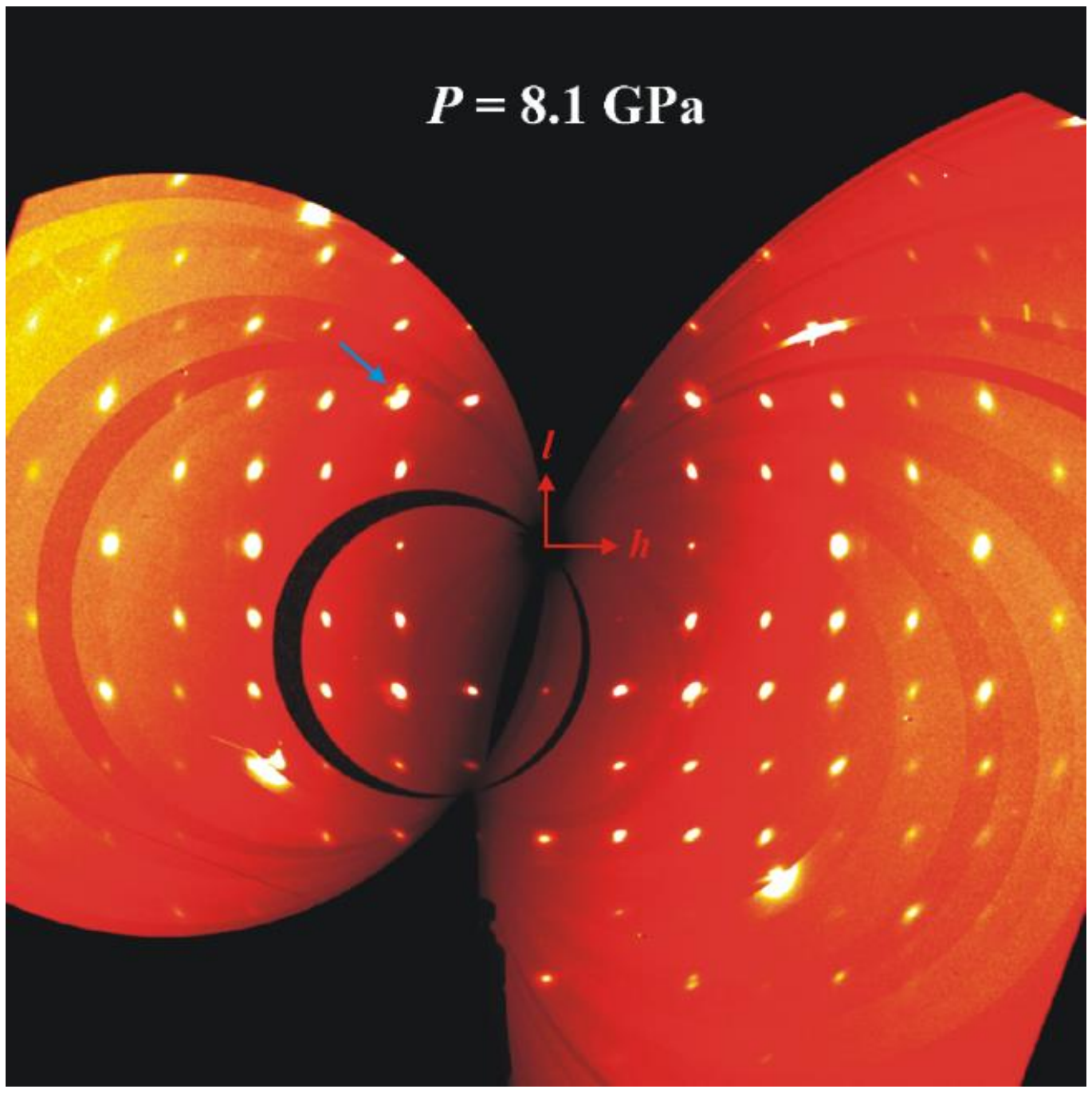

**Figure 3. Reconstruction of the *h*0*l* reciprocal layer of FeSe at 8.1 GPa and RT from single crystal XRD data.** Blue arrow marks a second *d*-segregated domain (see below).

The orthorhombic (*Pnma*) MnP-type structure is a slightly distorted hexagonal ($P6_3/mmc$) NiAs-type[33] that is evidenced, in the case of high-pressure FeSe, by its pseudo-hexagonal lattice ($c \approx \sqrt{3}\cdot b$). A similar conclusion could stem from a comparison of crystal geometry of the ambient-pressure tetragonal ($P4/nmm$) PbO-type structure of FeSe with the LT-LP orthorhombic one (*Cmma*): the latter is a slightly distorted variant of the former. Both distortions (from "tetra" ($P4/nmm$) to "ortho-I" (*Cmma*) and from "hexa" ($P6_3/mmc$)-to "ortho-II" (*Pnma*)) can be easily identified in symmetry terms:

(i) the phase transformation lowering FeSe crystal symmetry from $P4/nmm$ ($Z_p = 2$) to *Cmma* ($Z_p = 2$) is a proper ferroelastic transition, its mechanism consists of a macroscopic distortion of the tetragonal lattice which is induced by the non-diagonal strain tensor component $e_6 \equiv e_{xy}$ (irreducible representation $\Gamma_4^+ \equiv B_{2g}$ from the tetragonal Brillouin Zone (BZ) center)

(ii) displacements in the NiAs-type structure [$P6_3/mmc$ ($Z_p = 2$)] distorting it to the MnP-type [*Pnma* ($Z_p = 4$)] represent the eigenvectors for the $M_2^-$ phonon mode located at the hexagonal BZ face.

Due to the small magnitude of the above distortions we focus, for clarity, on the dominating mechanism transforming LP tetragonal (even pseudo-cubic) structure to the HP pseudo-hexagonal one. The non-direct relationship between LP and HP structures of FeSe underlines, first, the absence of a group-subgroup link between their space groups. However, this does not preclude a reconstructive phase transformation[34]. In this case one should expect non-negligible distortions of the structure, which indeed occur during the transformation (Fig. 4).

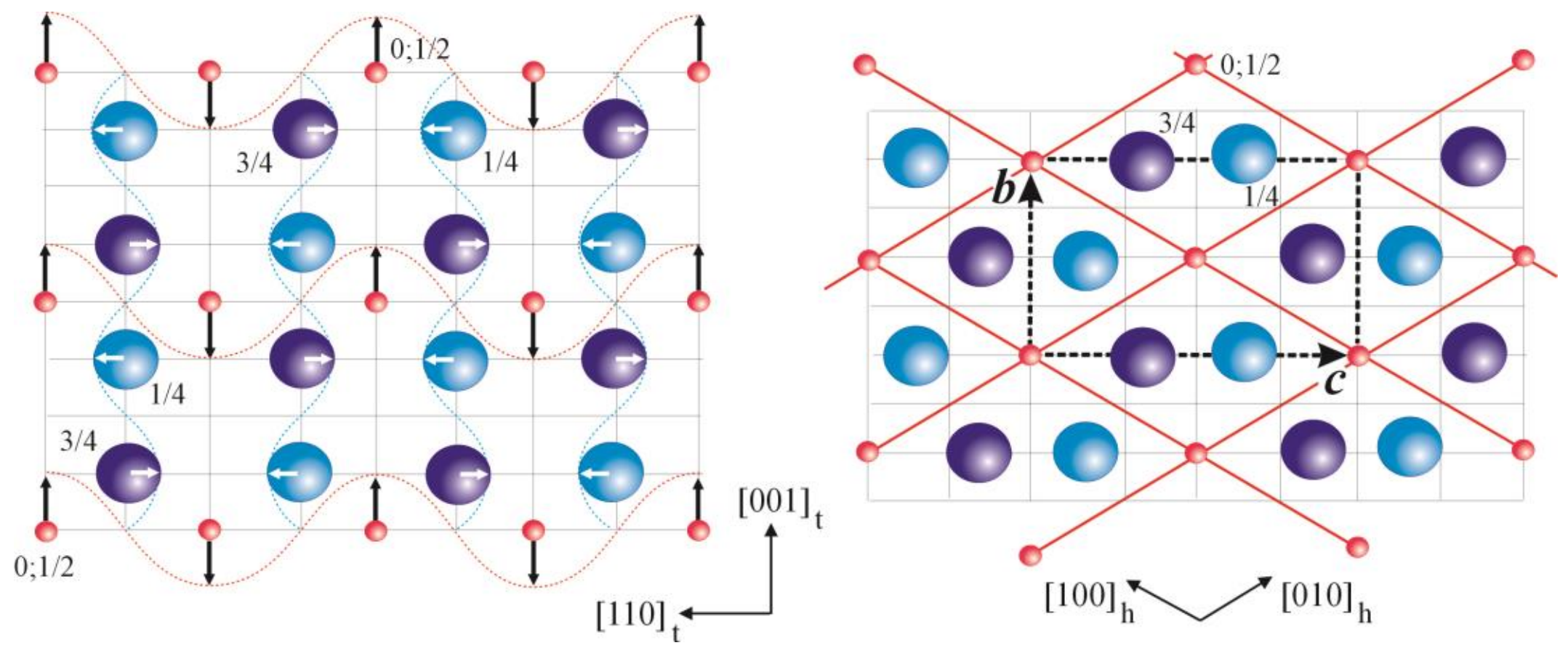

**Figure. 4. Transformation mechanism between LP (PbO-type) and HP (MnP-type) structures of FeSe.** Left panel: displacive $M_1^t$ phonon field applied to Fe (red spheres) and Se (blue spheres) sublattices. Right panel: result of combined structure distortion by the $M_1^t$ phonon field and macroscopic $e_{zz}^t$ strain. Red lines show hexagonal NiAs-type lattice; dashed lines mark the orthorhombic MnP-type structure cell.

We identify a primary order parameter (OP) that induces secondary (improper) macroscopic strain. Figure 4 illustrates the OP manifested as antiparallel displacements of both Fe and Se atoms. Despite the perpendicular directions, the two sets of the shifts belong to the same $M_1^t$ irreducible representation of the $P4/nmm$ group with the BZ-boundary k-vector $1/2(b_1^t + b_2^t)$. In addition to this critical phonon mode, the strain $e_{zz}^t$ (~0.3) is applied to the PbO-type parent structure making it metrically hexagonal. The reconstructive character of this transformation reflects its strong discontinuity and justifies, as a consequence, a *P-T* domain with two-phase coexistence. This is indeed observed in both our, and earlier, experiments[11,25,27,28]. It is worth noting that the above mechanism, and the corresponding epitaxial relations between LP and HP lattices, directs us to the text-book Burgers mechanism of the bcc-hcp transformation[34]; the difference here centers on the fact that the bcc-hcp transformation was suggested for mono-atomic close-packed structures while we deal here with the two-atoms derivative.

## High pressure (*P* > 6 GPa) behavior of FeSe and its implication on superconductivity

First, we note that the crystallinity of the HP FeSe phase undergoes a rapid microstructural degradation with pressure increase (for 10–12 GPa range at RT see Fig. 5S), even with He as the PTM. The degradation includes not only a rotational domain disorder (arching) but also a *d*-spacing dependent segregation (see also Fig. 3 for 8.1 GPa). The latter indicates that the application of pressure induces correlated deviations from the average structure, probably manifested as different local arrangements of the constituting chains of the face sharing $FeSe_6$ octahedra.

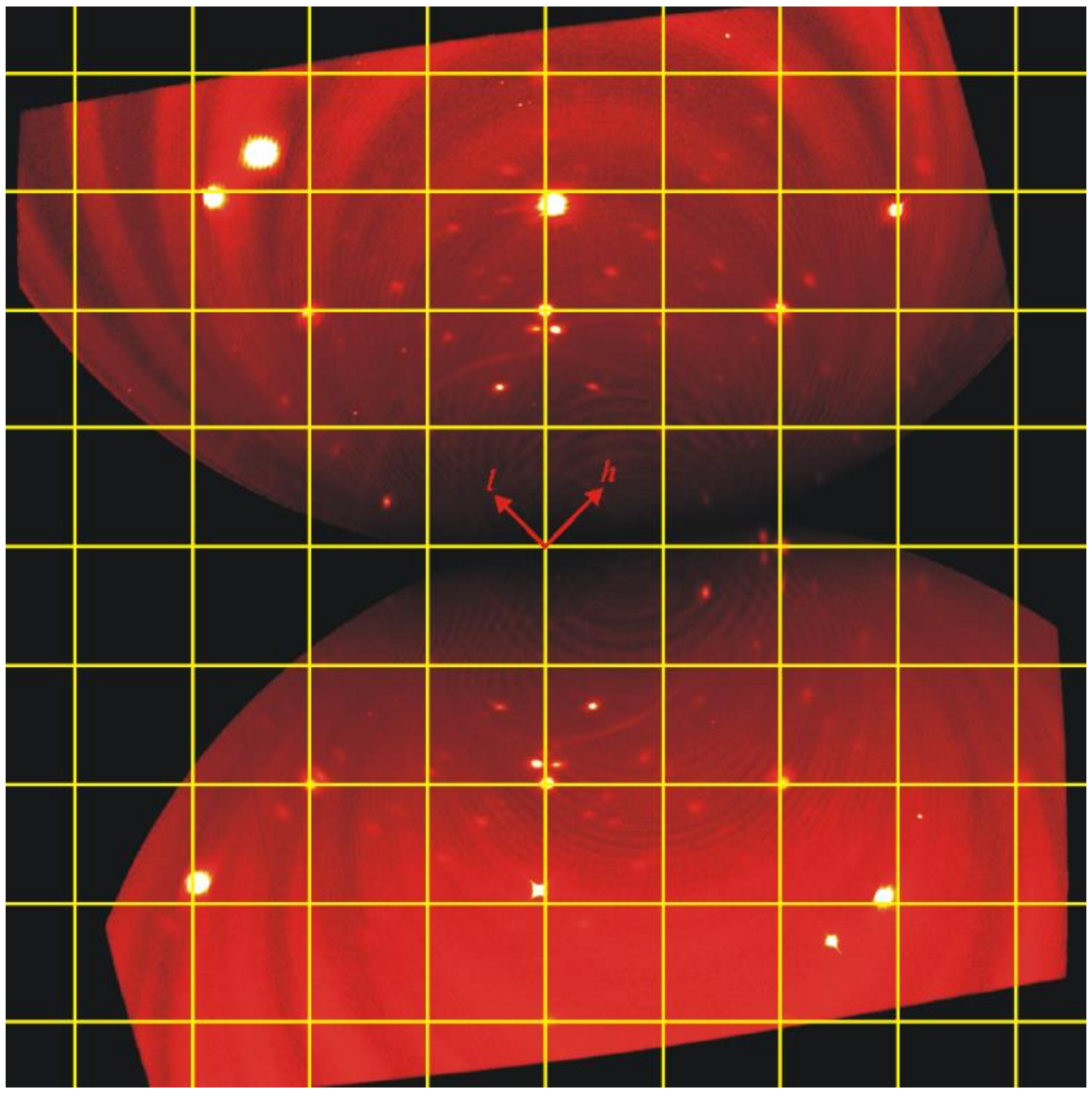

**Figure 5. Coexistence of the HP FeSe phase** (reciprocal lattice vectors indicated by red arrows) **and the LP FeSe *Cmma* phase** (*hk*0 unit cell grid in the tetragonal setting is represented

by yellow lines) **at 20 K and 8 GPa.** Very bright spots (eight in total) not commensurate with either FeSe lattice correspond to diamond reflections from the HP cell anvils.

At 20 K the HP *Pnma* FeSe phase appears between 6.2 and 6.9 GPa (Fig. 6S). It coexists (Fig. 5, 8 GPa) with the LP *P*4/*nmm* FeSe phase up to the highest measured pressure of 10 GPa where the HP *Pnma* form is dominant. At 50 K the HP *Pnma* is visible at 9.1 GPa (Fig. 7S) and, similarly, coexists with the LT FeSe phase (tetragonal) up to at least 14 GPa. At this temperature the HP *Pnma* phase likely starts to form at around 7 GPa, which corresponds to the interpolation point between the corresponding transition pressures at RT and 20 K (Fig. 2). However, it may not still be detectable with X-ray radiation due to the small phase fraction (the corresponding points are colored in cyan on the phase diagram, Fig. 2). Otherwise it would imply the formation of a "pocket" in the "ortho-II" phase around 50 K which would not be rational from a physical point of view.

The coexistence of the LP and HP modifications of FeSe in a wide pressure range of at least 4 GPa at 20-50 K confirms a first order character of the transformation with a relatively slow kinetics at these temperatures. Indeed, no coexistence of the LP and HP phases seems to occur at 300 K where the transformation takes place between 6.6 and 8.1 GPa (Figs. 8S, 3 and 9S,). The corresponding *P,T* phase diagram for FeSe constructed from our XRD analysis is compared to various available data of the literature in Fig. 2.

As demonstrated above, one important conclusion of our study is the proof that at low *T* the LP and HP phases of FeSe coexist above 6–7 GPa over a pressure range of 4 GPa at least (Figs. 5, 6S and 7S). However, at 20 K and 50 K we observe that the LP superconducting *Cmma* pseudo-tetragonal phase (and the AFM phase) is gradually squeezed out by its HP modification as shown by the relative intensities of Bragg spots measured at 20 K, presented in Fig. 6. At this point of the discussion, we should emphasize that a direct relationship between phase transformations seen by XRD data and independently measured superconducting properties is not easy for FeSe, since the LP *Cmma* – HP *Pnma* transition is first-order. In addition, it was shown for iron selenides that different experimental conditions, which include pressurization rates, *P-T* paths and PTMs, can dramatically influence transformations (see discussion above and ref. 31). The particular case of FeSe is delicate. For instance, in the recent work by Sun *et al.*[20], a superconducting transition was still observed above 7 GPa, with large diamagnetic signal at 7.8 and 8.8 GPa suggesting a bulk nature of superconductivity, although the temperature-dependent measurements (both resistivity and susceptibility) were performed after respective increases in pressure at RT. At this temperature, based on our study (Fig. 2), the formation of a pure non-superconducting HP *Pnma* phase is expected. However, because of the use of glycerol as a PTM by Sun *et al.*, which is less hydrostatic than the helium PTM used in the current work, and the multiple LT-HT cycling at lower pressures, the stability range of the LP superconducting *Cmma* was certainly enhanced so increasing the effective region of the two-phase coexistence. This is further confirmed by the report of Braithwaite *et al*[25] on an earlier generation of crystals grown in NaCl/KCl flux; using similar LT-HT cycling at each increased pressure step with argon as the PTM, the authors showed that superconductivity in FeSe survives up to 10.5 GPa and is fully destroyed above 12 GPa.

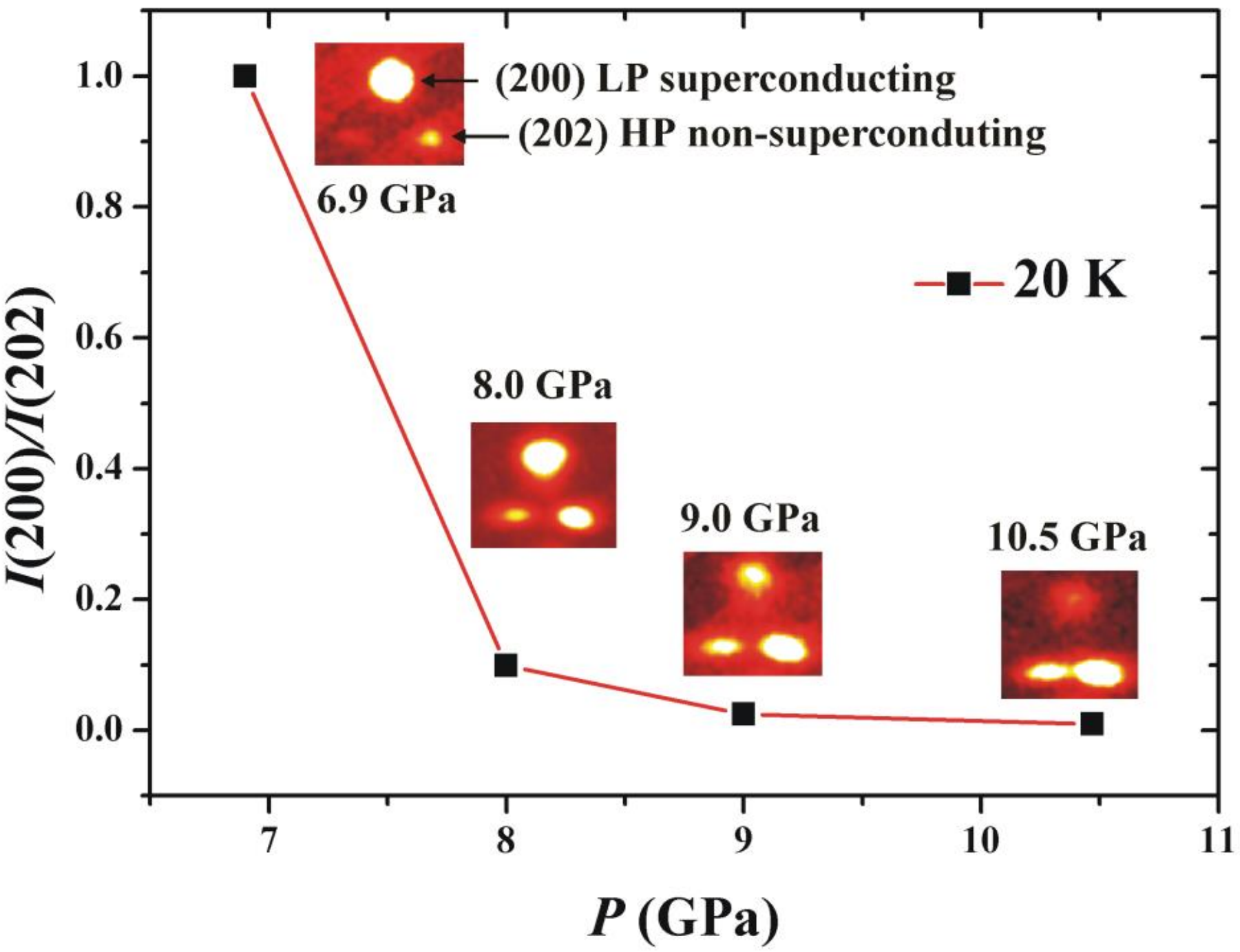

**Figure 6. Intensity ratio of the (200) and (202) peaks at 20 K of the LP and HP phases of FeSe, respectively, indicating a decrease in the concentration of the LP superconducting phase.**

Before the disappearance of superconductivity it is clearly established that the $T_c$ of FeSe first exhibits a plateau around 25 K in the 3–5 GPa range in polycrystalline samples, as concluded from magnetization and electrical resistivity measurements with the Daphne 7474 or Daphne 7373 oil as PTM[35], or at 20 K in the 2.5–5 GPa range in single crystals pressurized with glycerol as PTM[20]. Then follows a strong increase, reported in several studies, with a maximum $T_c$ around 6 GPa[36,20] (glycerol as PTM) or in the 6–8 GPa range[25,22] (argon gas as PTM). In all cases a decrease of $T_c$ is found by further pressurization above this critical pressure[11,20,35]. Above this pressure, all previous reports based on powder XRD show that the LP phase of FeSe becomes thermodynamically unstable and is replaced by its HP form[25,26,27]. As already stated, a direct comparison of structural and superconducting properties can only be obtained from diffraction experiments coupled to measurements of physical properties where the hydrostaticity conditions, *P-T* paths and rates would be identical. Such data do not exist for FeSe. Nevertheless, we remark that independently of the PTM used, the nature (polycrystalline or single crystal) and the exact stoichiometry of the sample used for structural or physical measurements, the appearance of the HP polymorph of FeSe and the $T_c$ decrease above 6–8 GPa is always observed. Therefore, these two properties are remarkably robust and can be correlated. In fact, the present single crystal XRD study shows clearly that the decreasing amount of the LP *Cmma* superconducting phase correlates with the opposite increasing amount of the HP phase (Fig. 6), exactly in the 6–10 GPa range where the decrease of $T_c$ has been measured[11,20,35] (Fig. 2, violet line). Hence the superconducting dome extends over two regions of the phase diagram: one corresponding to the pure pseudo-tetragonal

*Cmma* ("ortho-I") region and the second one to a mixture of the pseudo-tetragonal *Cmma* ("ortho-I") + HP orthorhombic ("ortho-II") phases.

We conclude that the initial LP increase in $T_c$ is an *intrinsic* effect of the LP "ortho-I" structure: pressure induced decrease of the distances between the FeSe sublayers implies an increase of the electron density in the superconducting planes and the associated modification of the spin fluctuations with a subsequent increase in $T_c$. The further decrease in $T_c$, and the full loss of superconductivity at higher pressures, results from the decrease in the *concentration* of the superconducting LP *Cmma* phase (composed of layers of edge-shared $FeSe_4$ tetrahedra) which is transformed into the HP *Pnma* phase (characterized by a 3D network of face sharing $FeSe_6$ octahedra). Above 6–8 GPa, the 2D tetragonal FeSe sublayers become uncorrelated and the effective distances between them increase resulting in a decrease of $T_c$ and the eventual loss of superconductivity. This indicates that superconductivity must be explained within a single mechanism based on superconducting 2D FeSe layers characteristic of the pseudo-tetragonal structure.

## Conclusions

With the help of X-ray synchrotron radiation and the use of high quality single crystals, the nature of the superconducting dome of the FeSe phase diagram is revealed, uncovering the relationship between topology and the macroscopic properties. Within the superconducting dome, we observe pressure- and temperature-dependent coexistence of two topologically different phases: a 2D-layered superconducting LP form and a 3D-framework HP form. Indeed, we demonstrate that pressurization above 6 GPa leads to a first order structural transformation of the superconducting phase to a non-superconducting one via nucleation and growth. We conclude that the change in the structural topology that, in turn, changes the local environment of Fe atoms in the HP *Pnma* FeSe polymorph is not favoring superconductivity. Thus, a key parameter for a further enhancement of superconducting properties in the FeSe system is a stabilization of the LP pseudo-tetragonal superconducting phase by controlled structural tunings, including intercalation and the formation of vacancies.

## Methods

Single crystals of FeSe studied in this work were grown at the Néel Institute (CNRS, Grenoble, France) using the chemical vapor transport (CVT) method based on a mixture of Fe, Se and $AlCl_3$/KCl chlorides heated in a two-zones furnace. The crystals were fully characterized at ambient pressure[37].

*P*-dependent single crystal XRD experiments were performed on the ID27 high-pressure beamline at the ESRF. Monochromatic radiation with wavelength 0.3738 Å was selected, and the data were recorded on a Perkin Elmer flat panel detector. To unambiguously conclude on all of the observed phenomena, a highly hydrostatic PTM has to be used[31]. Therefore He was chosen as the PTM due to its excellent hydrostatic properties up to at least 50 GPa[38]. To cover the pressure range to 20 GPa, diamond anvils with 500 µm culets and rhenium gaskets with 250 µm holes were used.

The pressure in the diamond anvil cell was controlled by an automatic pressure drive and was measured using the ruby fluorescence technique[39]. The pressure was changed from 0.1 GPa to a maximum of 19 GPa, with a typical step of 1 GPa at the temperatures of 20, 50 and 300 K. We used a custom He cryostat to achieve and control the low temperatures. The experimental diffraction patterns were influenced neither by gasket nor by the ruby spheres up to the highest studied pressures. The experimental single crystal data were integrated and corrected for absorption with the CrysalisPro package[40]. Structural solution was obtained with a help of SHELXT program[41]. The error bars are shown on the figures only if they are larger than the corresponding data symbols.

**Acknowledgements**

The authors thank to S. Bauchau (ESRF), J. Jacobs (ESRF) and P. Strobel (Néel) for technical support.

**Author contributions**

V.S., M.R., P.R., P.T. and M.M. conceptualized and designed the research; P.T. grew and characterized the FeSe single crystals; V.S., M.R., P.R., P.T. and M.M. conducted synchrotron X-ray diffraction experiments; V.S., M.R., P.R. and P.T. performed X-ray diffraction data analyses; V.D. and D.C. performed group-subgroup analysis; V.S., M.R., V.D., P.T. prepared figures; V.S. wrote the manuscript with the contributions from all the authors. All authors discussed the results and commented on the paper

**Competing financial interests**

The authors declare no competing financial interests

# Complex biphase nature of the superconducting dome of the FeSe phase diagram

V. Svitlyk[1]*, M. Raba[2,3], V. Dmitriev[4], P. Rodière[2,3], P. Toulemonde[2,3],

D. Chernyshov[4], M. Mezouar[1]

[1]ID27 High Pressure Beamline, European Synchrotron Radiation Facility, 38000, Grenoble, France

[2]Université Grenoble Alpes, Institut NEEL, F-38000 Grenoble, France

[3]CNRS, Institut NEEL, F-38000 Grenoble, France

[4]Swiss-Norwegian Beamlines, European Synchrotron Radiation Facility, 38000, Grenoble, France

*svitlyk@esrf.fr

## Supplementary information

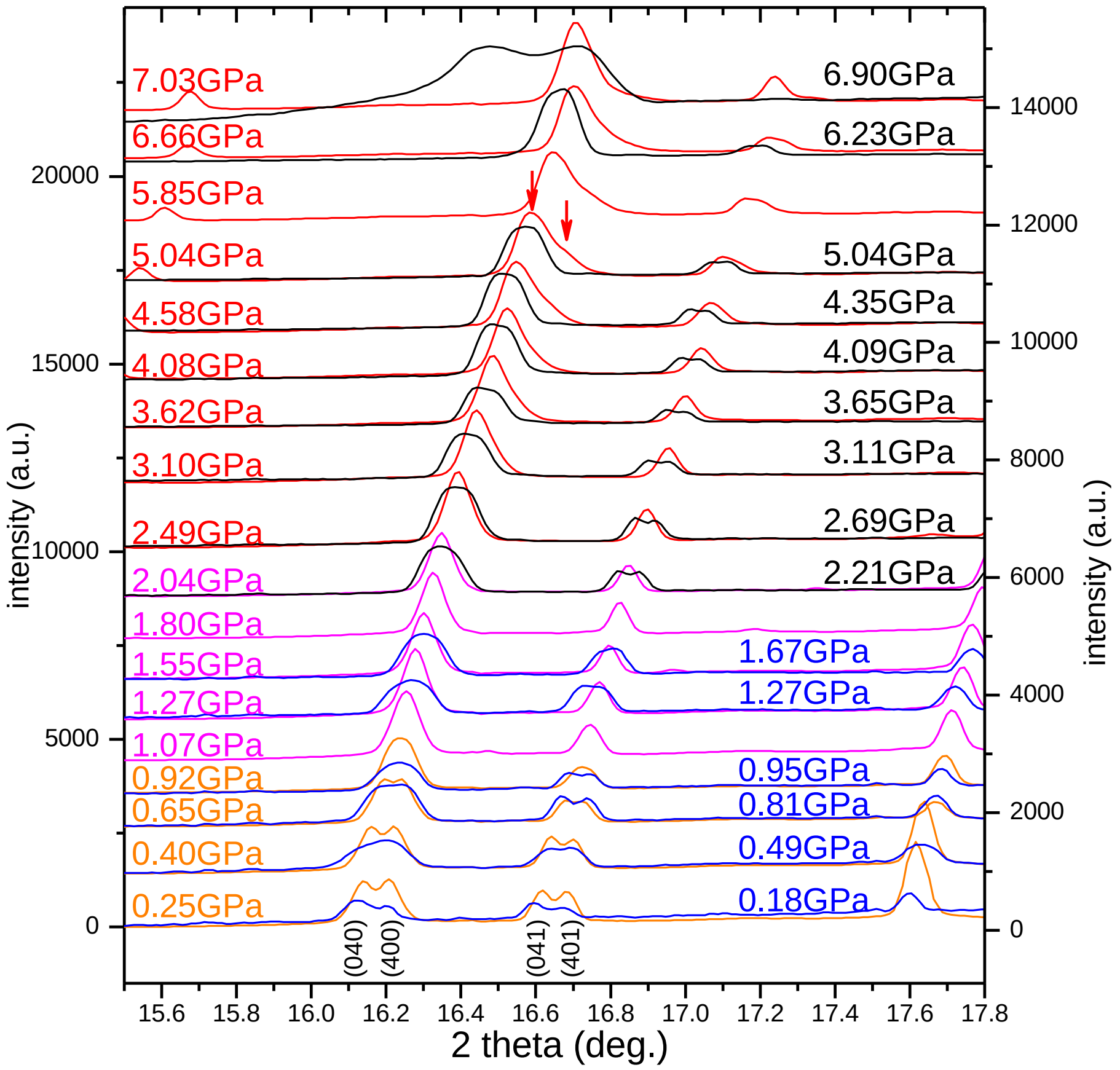

**Figure 1S. Comparison of integrated XRD patterns of the FeSe single crystal as a function of pressure at 50 K** (in yellow (*Cmma* “ortho-I” phase), magenta (*P*4/*nmm* tetragonal phase) and red (AFM phase) with pressures indicated on the left) **and at 20 K** (in blue (*Cmma* "ortho-I" phase) and black (AFM phase) with pressures indicated on the right).

As shown on Fig. 1S the splitting of (040)/(400) and (041)/(401) Bragg peaks of the initial orthorhombic *Cmma* phase is clearly always present at 20 K (black and blue patterns), showing that the distortion, at least an orthorhombic one, remains in the AFM phase, in agreement with the data published recently in the literature[1]. At 50 K this splitting disappear around 1 GPa, corresponding to the recovering of the tetragonal phase, and is again present above 2 GPa, i.e. in coincidence with the AFM phase detected by muons spectroscopy. The splitting is clearly observed at 4-5 GPa (see the double red vertical arrows) and measured as a broadening of the diffractions lines above 2 GPa. In addition, a careful inspection of other Bragg peaks, for instance the *hhl* ones (not shown here), reveals a possible supplementary splitting of them in the 2-6.6 GPa range, pointing towards a lower symmetry for the AFM phase, probably a monoclinic one. This point has to be confirmed by complementary experiments. For the series at 20 K, one observes already on the last XRD pattern acquired at 6.9 GPa the presence of the "ortho-II" *Pnma* phase (supplementary contribution around 16.5 deg.), not yet detected for similar pressure (7.03 GPa) at 50 K.

The lattice parameters calculated from the Bragg law at 20 K and 50 K are drawn on Fig. 2S and Fig. 3S and discussed in the main text of the article.

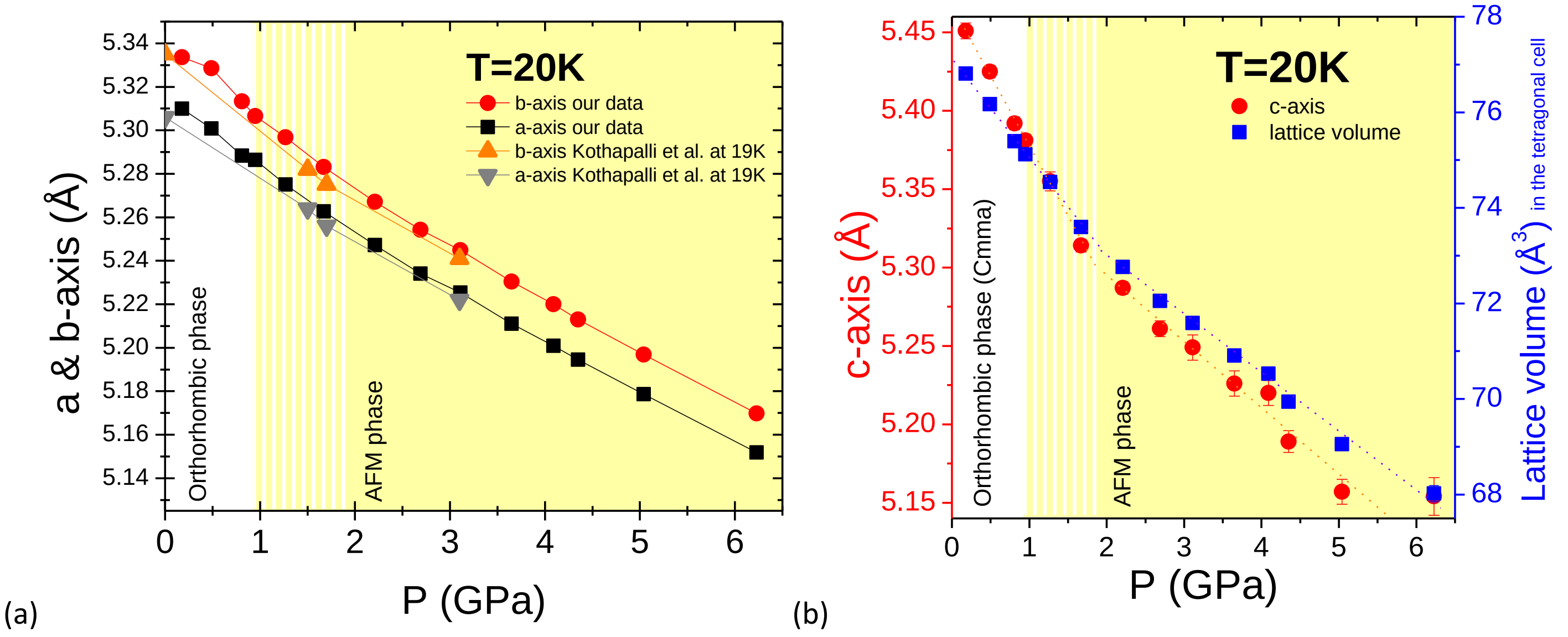

**Figure 2S. (a) *a*- and *b*-axis** (of the orthorhombic *Cmma* lattice) **pressure dependence at 20 K compared with the data of Kothapalli *et al.*[1]** (b) ***c*-axis and lattice volume** (calculated in the equivalent tetragonal cell) **as a function of pressure at 20 K**.

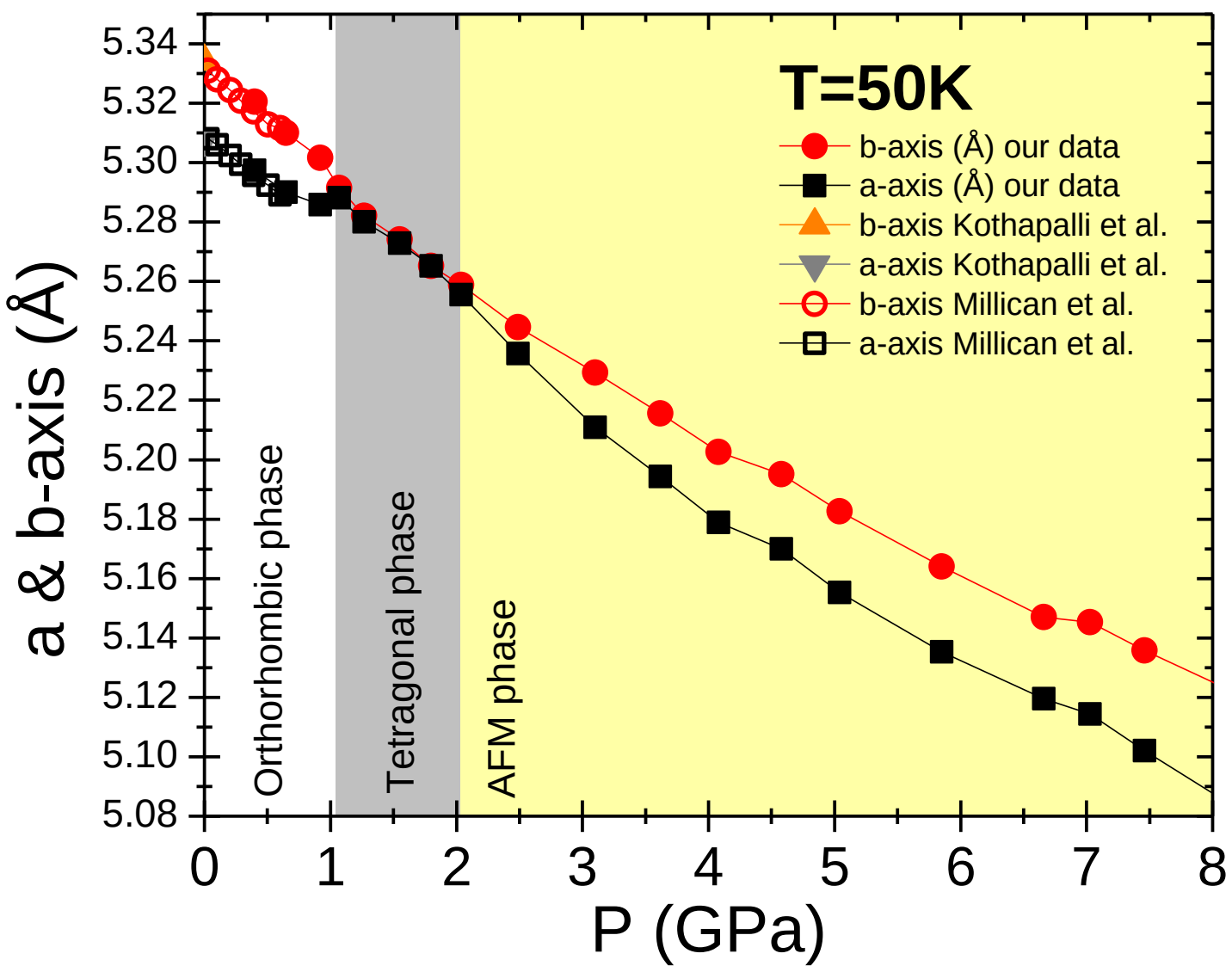

**Figure 3S.** ***a*- and *b*-axis** (of the orthorhombic *Cmma* lattice) **pressure dependence at 50 K compared with the data of Millican *et al.*[2] and Kothapalli *et al.*[1]**

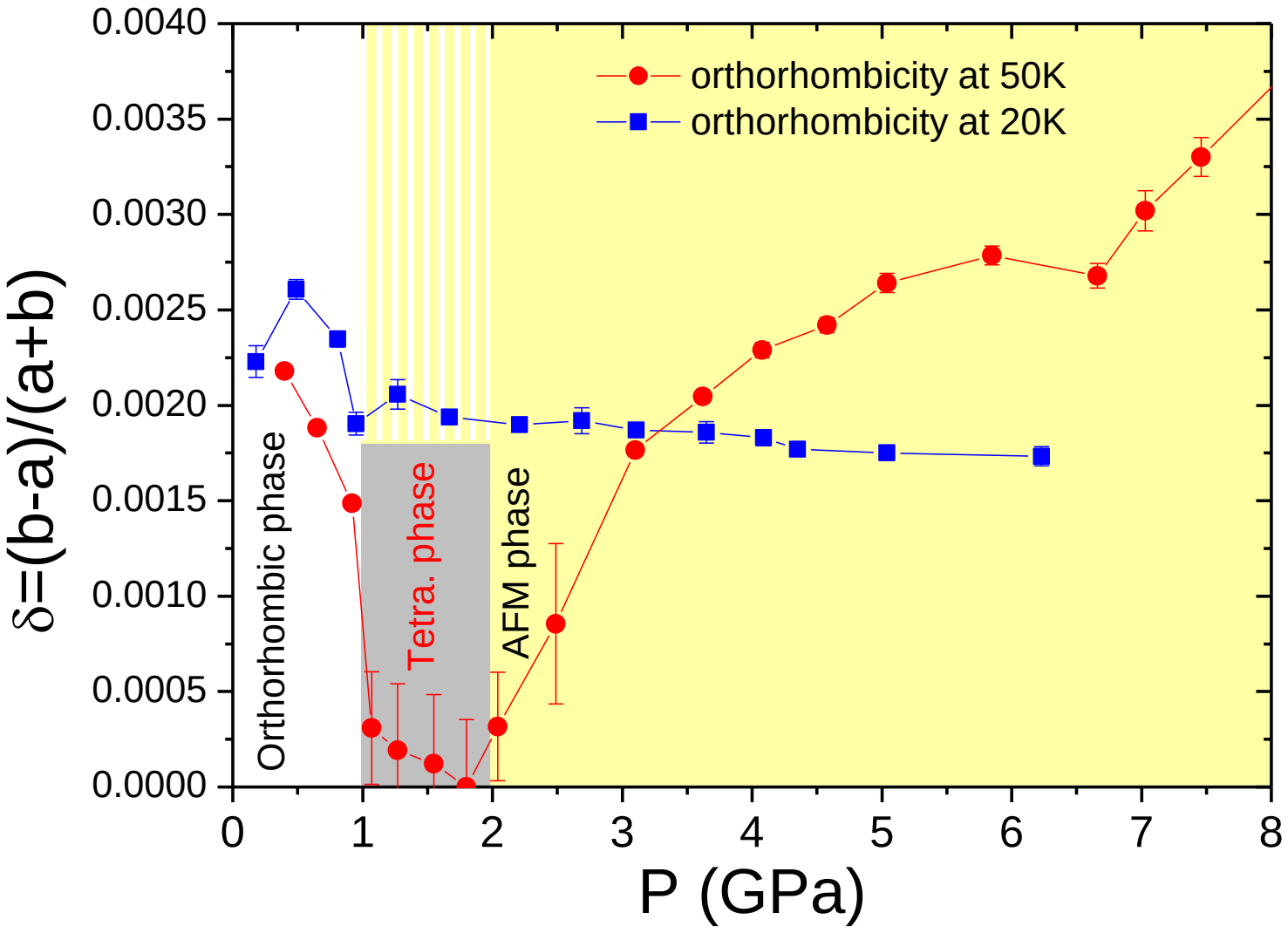

**Figure 4S. Orthorhombicity** (defined as $\delta = (b - a)/(b + a)$ parameters ratio) **of the ortho-I phase at superconducting, i.e. for P>1.7-2GPa,** (20 K, blue) **and non-superconducting, i.e. in particular in the AFM phase for P>2GPa,** (50 K, red) **temperatures.**

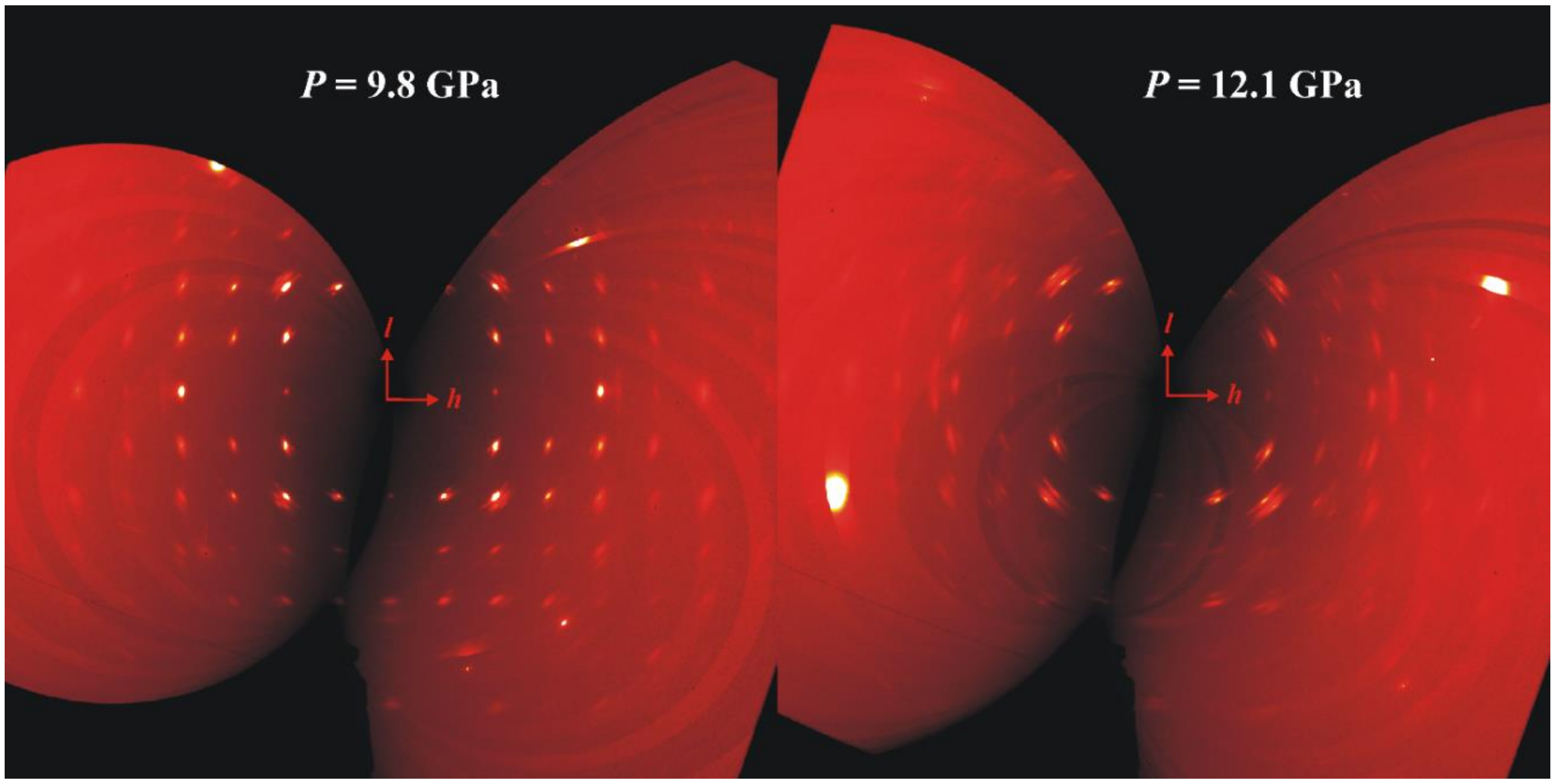

**Figure 5S. The *h0l* reciprocal layer of FeSe at 9.8** (left) **and 12.1 GPa** (right) **at 300 K illustrating a rapid degradation of the crystalline quality of the HP FeSe phase.**

The following figures with the slices of the reciprocal space of FeSe at 20, 50 and 300 K around the transition pressures illustrate the corresponding LP-HP phase boundaries on Fig. 2.

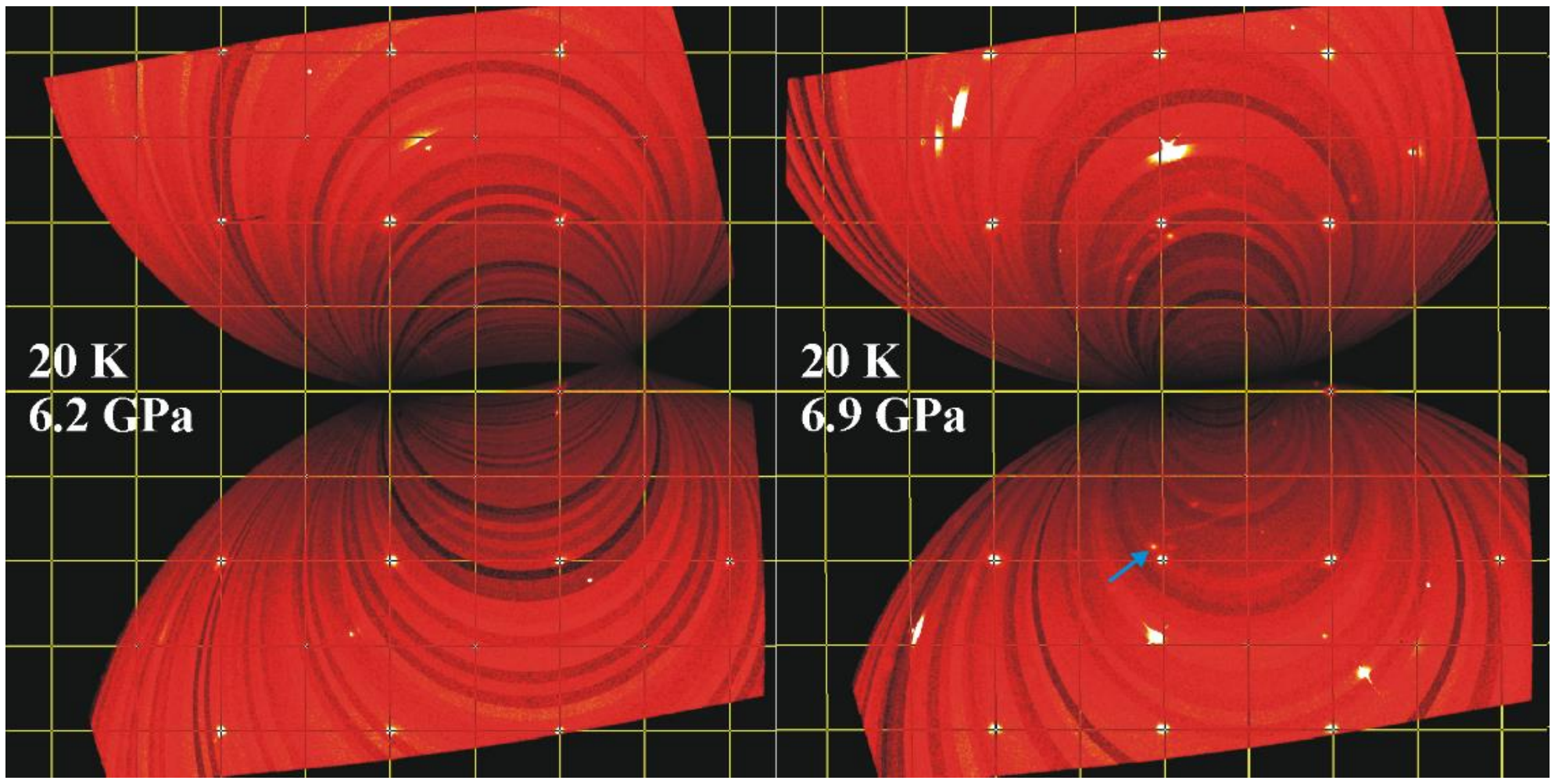

**Figure 6S. The *hk*0 reciprocal layers of FeSe at 6.2 and 6.9 GPa at 20 K. The blue arrow indicates the appearance of the HP *Pnma* phase** (here and afterwards the yellow grids correspond to the tetragonal/pseudo-tetragonal lattice of LP FeSe)

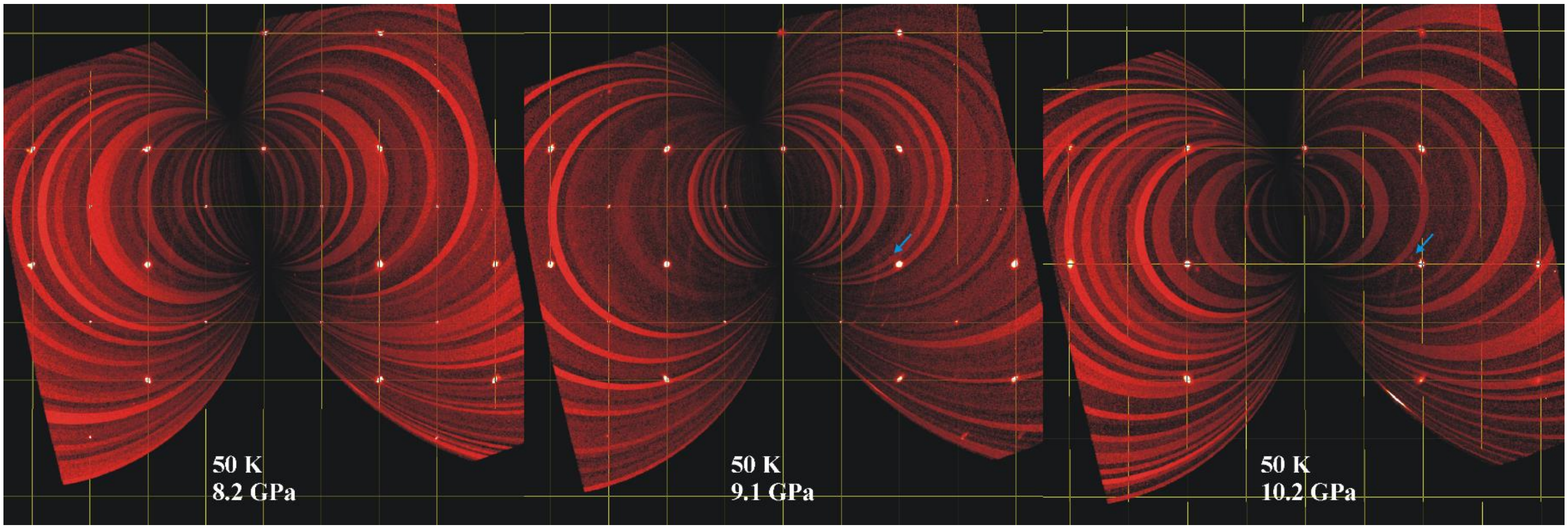

**Figure 7S. The *hk*0 reciprocal layers of FeSe at 8.2, 9.1 and 10.2 GPa at 50 K. The blue arrows indicate the appearance of the HP *Pnma* phase**

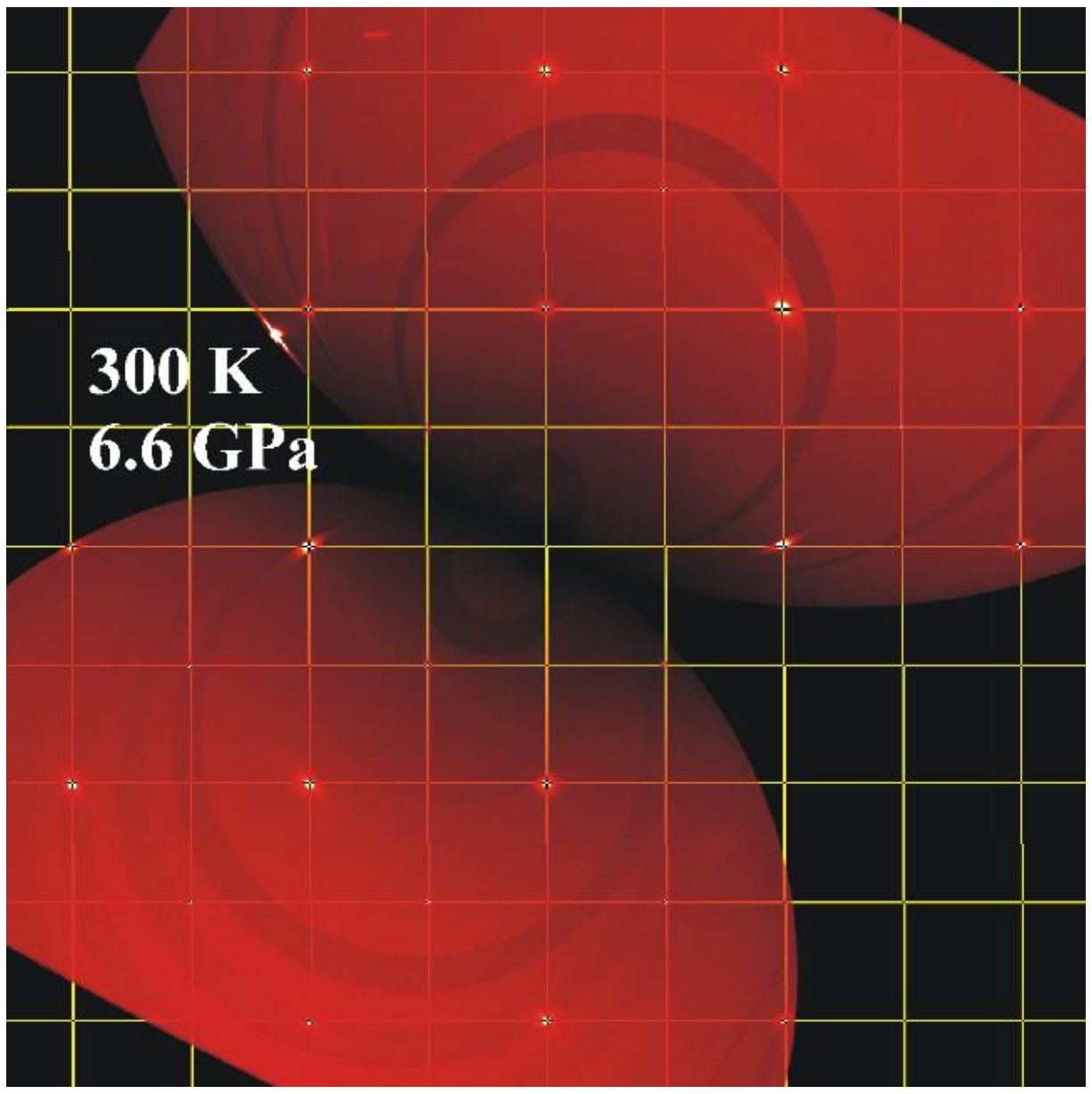

**Figure 8S. The *hk*0 reciprocal layer of FeSe at 6.6 GPa at 300 K before the transition to the HP *Pnma* modification. The next pressure point at 8.1 GPa corresponds to a single HP *Pnma* FeSe phase** (Fig. 3 in the main text)**.**

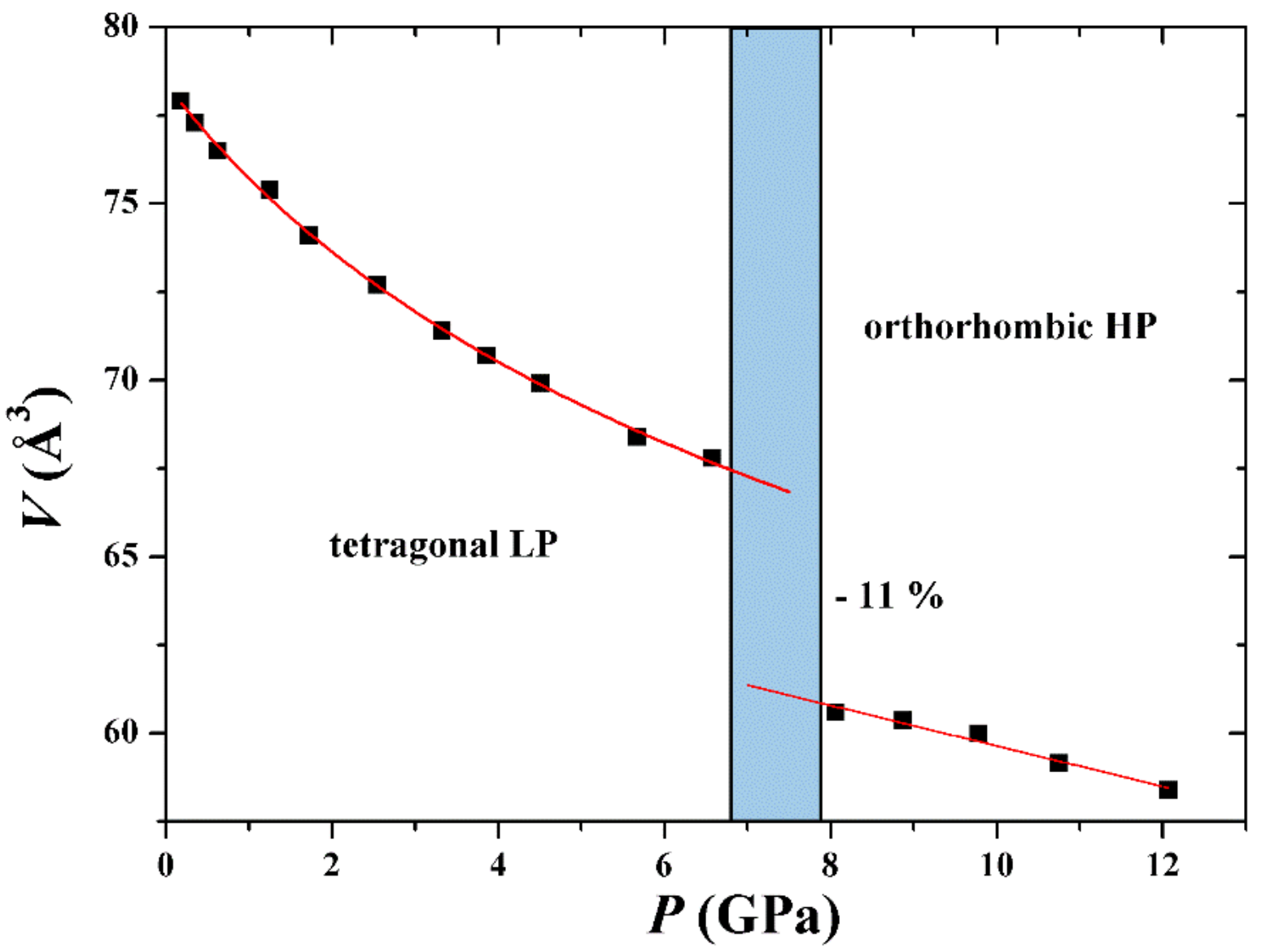

**Figure 9S. Phase transition and third order Birch-Murnaghan equation of state (EOS) of tetragonal FeSe at 300 K** ($V_0$ = 78.4(1), $B_0$ = 25(1), $B_0$’ = 7.1(7)). **A corresponding EOS for the high pressure orthorhombic FeSe modification have not been obtained due to the lack of reliable data points as a result of a rapid structural degradation at high pressure** (see the main text and Fig. 4S above).